\documentclass[12pt]{article}

\renewcommand{\baselinestretch}{1.3}
\normalsize

\usepackage{amsmath}
\usepackage{epsfig}
\usepackage[psamsfonts]{amsfonts}

\newcommand{\be}{\begin{equation}}
\newcommand{\ee}{\end{equation}}
\newcommand{\bea}{\begin{eqnarray}}
\newcommand{\eea}{\end{eqnarray}}
\newcommand{\no}{\noindent}
\newcommand{\3}{{\bf 3}}

\newcommand{\6}{{\bf 6}}

\begin{document} 
\renewcommand{\baselinestretch}{1.4}
\normalsize

\begin{center}
{\bf \Large N\normalsize O \Large R\normalsize ADIAL \Large E\normalsize XCITATIONS IN \Large L\normalsize OW \Large E\normalsize NERGY \Large QCD. II. 

\vskip .4cm

\vskip .4cm

{\bf \Large T\normalsize HE \Large S\normalsize HRINKING  \Large R\normalsize ADIUS OF  \Large H\normalsize ADRONS}}

\end{center}
\vskip .5cm

\large
\begin{center} {\tt Tamar Friedmann$^*$$^\dagger$}
\end{center}

\normalsize
 
\centerline{\sl Massachusetts Institute of Technology, Cambridge, MA 02139, USA}

\vskip .2cm

\centerline{\sl University of Rochester, Rochester, NY 14623, USA}

\vskip 1cm
\abstract{ 
We discuss the implications of 
our prior results  
obtained in our companion paper \cite{companionI}. 
Inescapably, they lead to three laws governing the size of hadrons, including in particular protons and neutrons that make up the bulk of ordinary matter: 
a) there are no radial excitations in low-energy QCD; b) the size of a hadron is largest in its ground state; c) the hadron's size shrinks when its  orbital excitation increases.   The second and third laws follow from the first law. It follows that the path from confinement to asymptotic freedom is a Regge trajectory. It also follows that the top quark is a free, albeit short-lived, quark. 

[For Note Added regarding experimental support, including the experiments studying muonic hydrogen, and other experiments, see last page.]}

\vskip 1cm

\vfill 
\no $^*$E-mail: tamarf at mit.edu 

\no $^\dagger$Current E-mail: tamarf at pas.rochester.edu

\thispagestyle{empty}
\newpage

\setcounter{page}{1}

Quarks and hadrons are known to be controlled by the strong interactions described by QCD. 
Even though it is well established that quarks  are the building blocks of hadrons, the quarks have never been seen in isolation (with the notable exception of the top quark, see below). This phenomenon of quarks, which is known as confinement and characterizes the strong interaction at low energies, is not yet well-understood and it is one of the major unsolved problems still facing particle physics. While many models and mechanisms for this phenomenon have been proposed,
  the actual dynamics of quarks at low energies are not yet known. 

In this paper, we rely on results obtained in our investigation of mesons \cite{companionI} to take a step towards a better understanding of the 
dynamics of the strong interactions, including confinement, asymptotic freedom, and the transition between them. In the process, we obtain a new set of relations between the size and energy of hadrons.

In \cite{companionI}, we developed an 
extended schematic model for hadrons, employing purely QCD ingredients with no extraneous input and no assumptions about interquark interactions. Our model extended the quark model to include certain diquarks as 
building blocks on equal footing with  quarks. The appropriate types of diquark building blocks were derived from hadron phenomenology, which led to flavor-antisymmetric diquarks for the meson spectrum, that is, the $\3_f$ of $SU(3)_f$ for the light mesons, the $\6_f$ of $SU(4)_f$ when the heavy charm is included, and so forth. For the baryon spectrum, the  appropriate diquarks were  color-antisymmetric, i.e. the $\3_c$ of $SU(3)_c$. 

Using this model, we first reclassified the 
meson spectrum 
into quark-antiquark and diquark-antidiquark states. The result 
inferred from this model was that 
all mesons that had been believed to be radially excited quark-antiquark states are actually orbitally excited diquark-antidiquark states;  no radial quantum number appeared in the classification of mesons. We then turned to the baryon spectrum and noted that the only baryons formerly believed to be radially excited -- namely the Roper resonances -- are actually made of two diquarks and an antiquark with orbital excitations. Hence, we observed that there was no radial quantum number in the classification of baryons either. 
Therefore, we were led to the conclusion that there are no radial excitations in the entire hadron spectrum. The result applies to all hadrons, including those that were previously considered "exotic" as they are part and parcel of the model 
as diquark antidiquark states
and are no longer exotic. The results also apply equally to the light and heavy hadron spectrum, thus providing a unified picture for the full hadronic spectrum. The recent discovery by the BELLE collaboration of the two heavy mesons $Z_b(10610)$ and $Z_b(10650)$, which were anticipated in our extended classification as diquark antidiquark states (table 3b of  \cite{companionI}; also see \cite{Karliner:2008rc}) appears to be a manifestation of our classification.

The above constitutes our first law:

\begin{quote}

{\textit {{The Law of  the Hadronic Spectrum}}}: {\it There are no radial excitations in low-energy QCD.}

\end{quote}

Now we shall discuss the implications of this first law.

By definition, whenever radial excitations between two particles do exist, the particles are pushed  apart. For example, a radially excited hydrogen atom has larger average distance between its proton and its electron (i.e. a larger radius) than the same atom in its ground state. As the radial excitation quantum number $n_r$ increases, so does the radius of the atom. Eventually, as $n_r\rightarrow \infty$, the radius becomes infinite and the electron is separated from the proton. This process is known as ionization of the atom.

It is therefore clear that the absence of radial excitations in the hadron spectrum is 
directly related to the prohibition on separation of the constituents of a hadron, 
that is, it is directly related to the phenomenon called quark confinement. 
In other words, there is no radial excitation in the presence of confinement. 
Since radial excitations are prohibited for hadrons, but other excitations -- such as orbital excitations -- are allowed, it must follow 
 that the distance between the quarks in excited states cannot be larger than their distance in the corresponding ground state, or else such excitations would have been prohibited for hadrons  just as radial excitations are.

Therefore, we now have:

\begin{quote}

{\textit {{The Law of Ground State Hadrons}}}: {\it The radius of a hadron is largest when the hadron is in its ground state.}

\end{quote}

\vskip .3cm
What can we say about the radius of an excited hadron? So far, we know only that it cannot be larger than the ground state radius. But does the radius stay the same or does it become smaller?

To answer this question, we first turn to the Particle Listings in the  PDG \cite{Beringer:1900zz} for data. 
The radii of only four hadrons ($\pi$, $K$, $p$, $\Sigma$) have been measured, and all four  are in their ground state. Lattice QCD calculations bring in but one more data point \cite{negele}, also for a hadron in its ground state ($\Delta$). These radii are displayed in Table 1, along with masses and densities. Another ground state hadron, the $\rho$ meson, arguably has a size similar to that of the pion \cite{Nussinov:2008}, though its size has not been measured 
(see also "note added", below). 

At first glance, the available data seems to tell us only about the radii of hadrons in their ground state, and nothing at all about radii of excited hadrons. However, we do not stop here. 

\begin{table}
\begin{center}
\small

{\bf Table 1: Measured sizes of ground state ($L=0$) hadrons}

\vskip .6cm
\begin{tabular}{|c|c|c|c||l|}
\hline
 \multicolumn{5}{|c|}{\mbox{{\bf Mesons}}  } \\ \hline
&Mass (MeV) &Radius (fm)&Density (g/cm$^3$) &Source\\ 
\hline \hline
$\pi ^\pm$&140&.672 &.20$\times 10^{15}$&PDG\cite{Beringer:1900zz}\\ \hline
$K^\pm$&494&.560 &1.2$\times 10^{15}$&PDG\cite{Beringer:1900zz}\\
\hline 
 \multicolumn{4}{c}{  } \\ \hline
 \multicolumn{5}{|c|}{\mbox{{\bf Baryons}}  } \\ \hline
 &Mass (MeV) &Radius (fm) &Density (g/cm$^3$) &Source\\ 
\hline \hline
$p$&938&.87 &.61$\times 10^{15}$&PDG \cite{Beringer:1900zz}\\ \hline
$\Sigma^-$&1197&.78 &1.1$\times 10^{15}$&PDG\cite{Beringer:1900zz} \\
\hline 
$\Delta$&1382&.650&2.1$\times 10^{15}$&Lattice\cite{negele}\\ 
&1425&.632&2.4$\times 10^{15}$&Lattice\cite{negele}\\
&1470&.614&2.7$\times 10^{15}$&Lattice\cite{negele}\\ \hline
\end{tabular}
\vskip .3cm
\normalsize

\end{center}
\end{table}

Recall that there is  a direct relation between the mass $m$ of a hadron and its orbital excitation quantum number $L$ given by the Regge trajectory equation \cite{Chew:1962eu,Wilczek:2004im}:
\be m^2=a+\sigma L \, ,\ee
where $m$ is the mass of the hadron, $a$ is an intercept that depends on the trajectory, and $\sigma$ is the slope. 
So an orbitally excited hadron ($L>0)$ is more massive than its corresponding ground state ($L=0)$. 

Now, if we inspect the hadronic masses displayed in Table 1, we find that for both mesons and baryons, radii are smaller when masses are larger: the $K^\pm$ is smaller than the $\pi ^\pm$, and the $\Delta$ is smaller than the $\Sigma ^-$ which is smaller than the $p$. 
It is in fact natural to associate  a higher mass with a smaller size -- for example, a Compton wavelength is inversely proportional to mass. 
It is also completely standard in physics to associate higher energies or large momenta with smaller distances, and this principle should apply to orbital excitations of a hadron.

So we have: 
\begin{quote}
{\textit {{The Law of Shrinking Radii}}}: 
{\it The radius of  a hadron decreases when the hadron's orbital excitation increases.}
\end{quote}

\vskip .3cm
We may express the Law of Shrinking Radii in the following way: 

\be \label{2nd3rd} {\Delta R \over \Delta L} < 0 ~,\ee
\vskip .4cm
\no where $R$ is the hadron's radius. 

\vskip .3cm
Before we turn to some implications of our laws, we shall compare them to properties of {\it atomic} radii. 

The radius of an atom as a function of its quantum numbers is well-known; it is given by:
 \be \label{atomicrad}
 <\hskip -.1cm R \hskip -.1cm >={a_0\over 2Z}[3n^2-L(L+1)]~, 
 \ee
 where $Z$, $a_0$, $n$, and $L$ denote the atomic number,  the Bohr radius,  
 the principal quantum number of the atom, 
 and the orbital quantum number of the atom, respectively. 

We see that precisely the opposite of the Law of Ground State Hadrons holds in atomic physics: for an atom, the ground state ($n=1$, $L=0$) is the $smallest$ state\footnote{Recall that $L<n$.} while for a hadron, the ground state is the $largest$ state.

Similarly, the opposite of the Law of Shrinking Radii holds in atomic physics. The variation of the atomic radius with $L$, keeping fixed $a_0$ and $Z$ as well as the radial quantum number $n_r=n-L-1$, is easily derived from equation (\ref{atomicrad}):

\be \label{delraddelL} 
{\Delta \hskip -.1cm <\hskip -.1cm R\hskip -.1cm>\over \Delta L}= {a_0\over 2Z}(6n-2L-1) > 0.
\ee
This means that when the radial quantum number (and the number of radial nodes) is held fixed, the radius of an atom is larger when its orbital angular momentum is higher. (Compare to equation (\ref{2nd3rd}) for hadrons.)\footnote{The author is grateful to Guy de Teramond for discussions of this point.}

In retrospect, it is natural to expect fundamental differences between hadronic and atomic radial properties even if only because confinement of hadrons and ionization of atoms are opposite phenomena that are fundamental to their respective systems.

\vskip .5cm
\noindent Now we shall turn to implications of the three laws.

\vskip .3cm
\no \underline{{\tt Size-energy relation }}  Our investigation has uncovered a new set of relations between two fundamental properties of hadrons: their size and their energy. The higher the energy level, the smaller the size of the hadron. This set of relations is a new QCD effect.

\vskip .3cm
\no \underline{{\tt The path from confinement to asymptotic freedom }} As is well-known, the QCD coupling decreases at high energies and short distances, a phenomenon known as asymptotic freedom. Conversely, the coupling increases with decreasing energy, becoming large at low energies and long distances; the low-energy regime is characterized by  confinement whereby quarks are bound together as hadrons. 

We have shown that as the hadron's orbital excitation quantum number $L$ gets larger and larger, the radius of the hadron gets smaller and smaller. 
At some critical stage in this process, the radius is so small, the energy so high, and the coupling so weak that we have entered the regime of asymptotic freedom: the quarks become free and the hadron loses its structure.

Recalling that a series of hadrons in which each successive hadron has one more unit of orbital angular momentum $L$  is named a Regge trajectory \cite{Chew:1962eu,Wilczek:2004im,Selem:2006nd,companionI}, we 
have the following corollary of the Law of Shrinking Radii:

\begin{quote}
{\textit {{Corollary}}}: {\it The path from confinement to asymptotic freedom is a Regge trajectory.}
\end{quote}

Each Regge trajectory terminates at some critical value $L_c$ of $L$; above $L_c$, the quarks become free and they are not bound as hadrons. The number of known hadrons in a  trajectory \cite{companionI,Selem:2006nd}, which ranges between 3 and 6, sets a lower bound on the value of $L_c$ for each trajectory.

The process of decreasing radius, increasing $L$, and reaching asymptotic freedom is an explicit manifestation of 
the concept of antiscreening which is so fundamental to QCD \cite{Gross:2005kv,Gross:1973id}: the smaller the distance between the quarks, the smaller the effective color charge of one quark as seen by another, and the weaker the interaction between them.

The process is also a manifestation of a principle first put forth by Collins and Perry \cite{Collins:1974ky}, who explained that at sufficiently high densities, 
matter consists of a soup of asymptotically free quarks (and gluons). In our work, as $L$ gets larger,
 the hadron's mass gets larger and its radius smaller, so the density of the quarks in the hadron is high. Simultaneously, the QCD coupling, strongest when the hadron is in its ground state, becomes weaker and weaker as $L$ gets larger, 
  so the quarks become free. Therefore, asymptotic freedom and high density naturally go hand in hand. 

It follows that if this process can be carried out  for a large number of hadrons simultaneously, it could produce the quark-gluon plasma (QGP). So far, the QGP has been searched for and possibly produced only through heavy ion collisions \cite{nucl-ex/0603003}.

\vskip .3cm
\no \underline{{\tt No elongated flux tubes }} 
The commonly held assumption of many QCD models is that 
orbital excitations cause a hadron's size to increase. This assumption appears in many forms: 
  in the bag model, string-like solutions of the bag with large angular momentum are assumed to have an  elongated shape \cite{Johnson:1975sg}; 
  in flux-tube or string models, the flux tube or string is elongated at large $L$ 
   \cite{IsgurPatonI,Nambu:1974zg,Barnes:1995hc,Shifman:2007xn,Glozman:2004ww}; 
   the flux tube is also assumed to have {\it minimum} length of 1fm \cite{IsgurPatonI}; in potential models, the size of excited hadrons is increased \cite{Badalian:2002xy}; 
  and in many, if not all, models, it is assumed that when $L>0$ there is  a "centrifugal barrier" that pushes the quarks apart.  This assumption is intuitively convincing. But our results, while counter-intuitive, rely on no assumptions other than allowing quarks and diquarks to be building blocks for hadrons. Hence our results are model-independent. 
They imply that, remarkably, QCD overcomes the centrifugal barrier\footnote{Phrase graciously provided by Frank Wilczek.} so that when a hadron is orbitally excited, the quarks come closer to each other instead of being pushed apart.
 
 Our  laws  are also consistent with the model-independent results of lattice QCD. There, it has been shown that a color string actually breaks in lieu of stretching beyond around 1fm \cite{Bali:2005fu}; indeed, our  laws together with the measured radii displayed in Table 1 show that the radius of a hadron never exceeds around 1fm: it is around 1fm in its ground state and shrinks for all excited states. 

Further, our results are consistent with several experimental observations; see "note added" below.

\vskip .3cm
\no \underline{{\tt It is lonely at the top}}  The top quark is the only quark which has been observed on its own, i.e. not within a hadron.  It is also the only quark which has {\it never} been observed within a hadron --  there are no top mesons or top baryons. The top quark was first observed in pairs in 1995 \cite{topCDF, topD0,Beringer:1900zz} and more recently has been produced singly \cite{Abazov:2009ii}. It has mass over 170 GeV and a very short lifetime. 
 
 It has been standard to interpret the top quark's behavior by saying that it "decays before hadronizing" \cite{Beringer:1900zz,Bigi:1986jk}. We suggest a different interpretation that arises naturally from the rest of our results:  the top quark is so massive that it is already at such high energy and density that it lives in the asymptotically free regime where there is no confinement -- and no hadrons. It is a free, albeit  short-lived, quark.

\vskip .3cm
\no \underline{{\tt Ordinary matter}} The three laws apply to all hadrons, so in particular they apply to protons and neutrons. The protons and neutrons are the constituents of nuclei which make up almost the entire mass of the ordinary matter that surrounds us. The fact that their size is maximal in their ground state  and shrinks when they are excited should therefore have potentially significant ramifications for  the properties of all ordinary matter.

\vskip .3cm
\no \underline{{\tt Epilogue}} We set out to investigate the hadron spectrum by extending the quark model to include diquarks; we introduced no extraneous input or assumptions, and we employed only purely QCD ingredients: quarks and diquarks as building blocks for hadrons. In the resulting classification of the spectrum, all known hadrons emerged as  combinations of quarks and diquarks, with at most orbital excitations; there was no radial quantum number. Further, no hadrons were left out of the classification as "exotic."  This model led to the predictions of new particles as well as the introduction of isorons (iso-hadrons) and magic quantum numbers which appear to be the quantum numbers of low-lying glueballs expected in lattice QCD.
 In the process, a new set of relations between two fundamental properties of hadrons  - their size and their energy -  was uncovered. Inevitably, three laws governing the size of hadrons were put forth, providing a novel explanation for the transition between low energy and confinement on the one hand, and high energy and asymptotic freedom  -- where the top quark resides --  on the other hand. A new QCD effect was introduced whereby a hadron shrinks. The results obtained apply equally to light and heavy hadrons.

These results constitute simple and testable predictions about a fundamental property of hadrons: their size.

\vskip .3cm
\no \underline{{\tt Note added}}: Nine months after this paper was originally posted to arXiv \cite{arxiv}, an experiment studying muonic hydrogen \cite{pohl}, repeated more recently \cite{antognini}, observed a smaller size of the proton than previously expected, consistent with our predictions. It is possible that this is a manifestation of our three laws, and may be a QCD, rather than QED, effect. Further experimental results consistent with our predictions  appeared in  \cite{rhosize}, where it was reported that the HERMES experiment found shrinkage of the $\rho$ meson, and in \cite{Belle:2011aa}, where charged bottomonium-like "exotic" states, the $Z_b(10610)$ and $Z_b(10650)$, were discovered by the BELLE collaboration. The latter fit nicely in our classification tables of \cite{companionI} as our predicted isovector made of a diquark and an antidiquark. 

Further experimental verification of our results could come from comparing measurements of the radii of hadrons at different excitation levels, perhaps a measurement of all radii of a specific Regge trajectory such as any of the trajectories listed in \cite{companionI,Selem:2006nd}.

\vskip 1cm
\no {\bf Acknowledgements:} 
I am grateful to Frank Wilczek, who gave me a glimpse into his work on baryon systematics, and in response to my question "what about mesons?" encouraged me to pursue them. This work is the result. 
I am also grateful to Robert L. Jaffe, Howard Georgi, Richard Brower, Usha Mallik, Hulya Guler, Dan Pirjol, Ayana Holloway, and Guy de Teramond  for helpful discussions.
This work was supported in part by funds provided by the U.S. Department
of Energy (DOE) under cooperative research agreement DE-FC02-94ER40818 and in part by US DOE Grant number DE-FG02-91ER40685.

\end{document}